\documentclass[letterpaper, 10pt, conference]{ieeeconf}

\IEEEoverridecommandlockouts 
\usepackage{cite}
\usepackage{amsmath,amssymb,amsfonts}
\usepackage{amsmath}
\usepackage{algorithmic}
\usepackage{graphicx}
\usepackage{hyperref} 
\usepackage{textcomp}
\usepackage{xcolor}
\usepackage[normalem]{ulem}
\usepackage{url}

\definecolor{Aqua}{RGB}{131,211,212} 
\definecolor{Teal}{RGB}{45,129,131} 
\definecolor{DarkTeal}{RGB}{5,112,145} 
\definecolor{DarkRed}{RGB}{145,12,7} 
\definecolor{Orange}{RGB}{244,129,83} 
\definecolor{Gold}{RGB}{219,172,52} 
\definecolor{Orange}{RGB}{110,54,42} 
\definecolor{Rose}{RGB}{255,86,87} 
\definecolor{BlueGrey}{RGB}{99,143,169} 
\definecolor{Red}{RGB}{224,69,86} 

\hypersetup{colorlinks=true,linkcolor=purple,filecolor=mnodea,urlcolor=cyan,citecolor=blue}

\newenvironment{list4}{
  \begin{list}{$\bullet$}{
      \setlength{\itemsep}{0.05cm}
      \setlength{\labelsep}{0.2cm}
      \setlength{\labelwidth}{0.3cm}
      \setlength{\parsep}{0in}
      \setlength{\parskip}{0in}
      \setlength{\topsep}{0in}
      \setlength{\partopsep}{0in}
      \setlength{\leftmargin}{0.21in}}}
      {\end{list}}

\newcommand{\TC}[1]{\textcolor{black}{{#1}}}

\newcommand{\EA}[1]{\textcolor{black}{{#1}}}
\newcommand{\ND}[1]{\textcolor{black}{{#1}}}

\title{\huge \bf Experimentally-Driven Analysis of Stability in Connected Vehicle Platooning: Insights and Control Strategies}

\author{Niladri Dutta$^1$, Elham Abolfazli$^1$, and Themistoklis Charalambous$^{1,2,3}$
\thanks{$^1$Department of Electrical Engineering and Automation, School of Electrical Engineering, Aalto University, Espoo, Finland. E-mails: {\footnotesize
name.surname@aalto.fi.}
}
\thanks{$^2$Department of Electrical and Computer Engineering, School of Engineering, University of Cyprus, Nicosia, Cyprus.}
\thanks{$^3$FinEst Centre for Smart Cities, Tallinn, Estonia.}
\thanks{This work has been partly supported by the European Commission through the H2020 project Finest Twins (Grant agreement No. 856602) and the H2022 European Research Council (ERC) project MINERVA (Grant agreement No. 101044629).}
}

\begin{document}
\maketitle

\begin{abstract}
This paper presents the development of a tangible platform for demonstrating the practical implementation of cooperative adaptive cruise control (CACC) systems, an enhancement to the standard adaptive cruise control (ACC) concept by means of Vehicle-to-Everything (V2X) communication. It involves a detailed examination of existing longitudinal controllers and their performance in homogeneous vehicle platoons. Moreover, extensive tests are conducted using multiple autonomous experimental vehicle platform topologies to verify the effectiveness of the controller. The outcomes from both simulations and field tests affirm the substantial benefits of the proposed CACC platooning approach in longitudinal vehicle platooning scenarios. This research is crucial due to a notable gap in the existing literature; while numerous studies focus on simulated vehicle platooning systems, there is lack of research demonstrating these controllers on physical vehicle systems or robot platforms. This paper seeks to fill this gap by providing a practical demonstration of CACC systems in action, showcasing their potential for real-world application in intelligent transportation systems.


\end{abstract}


\section{Introduction}
\label{sec:intro}



The growing dependence on cars has resulted in a large number of vehicles on the road, placing a significant strain on the road infrastructure and raising the risk of accidents and traffic congestion. Research nowadays focuses on automotive system technology for providing intelligence to transportation systems in order to enhance traffic flow, road safety, and efficiency. Towards this end, one approach is to utilize wireless inter-vehicle communication and onboard sensors (found in modern cars) to facilitate vehicle collaboration.

The formation of vehicle platoons by means of CACC systems is one of the compelling applications that is now gaining even more popularity \cite{AlAlam2010AnPlatooning,Alam2014Fuel-efficientPlatooning,Alam2015Heavy-dutyEfficiency,Bekiaris2023_TITS, Bekiaris2024_IJRNC, Bekiaris2024_TIV}. A vehicle platoon refers to a group of closely linked vehicles traveling in close proximity, with each vehicle maintaining specific states like position, velocity, and acceleration through measurement sensors. 
CACC stands as a promising solution for enhancing intelligent mobility. By enhancing the ACC algorithm, this technology enables vehicles to share their system state information via wireless communication with other vehicles in the platoon, allowing them to travel with reduced inter-vehicular distances from their preceding vehicles while matching their speed and maneuvers. The lead vehicle can either be operated by a human or it can be an autonomous vehicle following a designated trajectory, potentially making the entire platoon fully autonomous.  This offers the advantage of enhancing traffic flow without necessitating additional road infrastructure. Through minimizing the overall aerodynamic drag in platoon vehicles and optimizing their individual speeds, CACC systems also lead to decreased fuel consumption and subsequently reduced carbon emissions \cite{Lammert2014EffectMass}.

Since the vehicles are interconnected within the platoon, this combines all their subsystem dynamics into a unified system. This necessitates an understanding not only of an individual vehicle's dynamics but also the dynamic properties of the vehicles \cite{deGeus2021PracticallyApproach}.
Vehicles in a platoon utilize control algorithms to set velocity and direction, ensuring safe trajectories. This work addresses longitudinal controller design for a vehicle platoon, considering lateral guidance control separately \cite{Chien1992AutomaticVehicle-following,Eyre1998AVehicles,Eyre1997StringControllers,Rajamani2002Semi-autonomousSystems,Yanakiev1996AAHS}, forming a vehicle platoon when applied simultaneously.
Numerous research works have showcased the application of a well-designed control system for various predecessor leader following (PLF) systems in theory and in ideal simulation settings \cite{Chien1992AutomaticVehicle-following,Eyre1998AVehicles,Eyre1997StringControllers,Rajamani2002Semi-autonomousSystems,Yanakiev1996AAHS}. However, there has been limited work on the exploration of these controllers in physical environments with hardware components, with only a few exceptions \cite{ma2013hybrid,landolfi2022delay,diab2010testing,rajamani2000design}.

The aim of this work is to investigate the efficacy of distributed controllers on a robotic vehicle platoon platform that simulates a real-world situation. For doing so, we have chosen to implement the controller developed in \cite{Abolfazli2021OnDelays}, which takes into consideration communication delays among vehicles. The goal is to investigate \emph{i)} the impact of communication limitations (bandwidth constraints, delays, packet losses), \emph{ii)} uncertainties (measurement and model uncertainties) on homogeneous platoons, and \emph{iii)} to ascertain if the controller proposed in \cite{Abolfazli2021OnDelays} can maintain overall system stability for a platoon leader with varying velocity over time. 




The remainder of this paper is organized as follows. Section~\ref{sec:preliminaries} provides some preliminaries on vehicle platoons.
Section~\ref{sec:problem_description} presents the vehicle model, the spacing policy and the platooning controller with the internal stability and string stability. The system architecture of the experimental platform is explained in Section~\ref{sec:system_architecture}. Section~\ref{sec:experiments} presents the experimental setup and its results followed by an analysis of the results obtained. Finally, in Section~\ref{sec:conclusions_futuredirections} we conclude the paper and discuss future directions.

\section{\TC{Preliminaries}}
\label{sec:preliminaries}



\subsection{\TC{Categories of platoons based on vehicle types}}

\begin{figure}[b]
    \centering
    \includegraphics[width=0.49\textwidth]{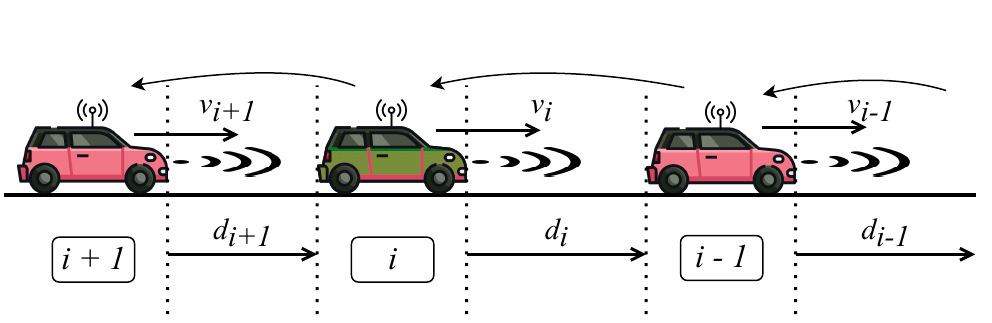}
    \caption{A part of the vehicle platoon used for introducing the notation. 
    }
    \label{fig:platooning_notation}
\end{figure}

Based on the types of vehicles in a platoon, there are two categories - homogeneous vehicle platoons and heterogeneous vehicle platoons. A platoon is considered homogeneous when all its vehicles share identical dynamics and physical constraints. Conversely, if the platoon comprises non-identical vehicles, it's categorized as a heterogeneous platoon. In this work, all experiments focus on homogeneous platoons, where each vehicle is equipped with the same controllers, sensors, and communication systems to facilitate inter-vehicular cooperation \cite{Eyre1998AVehicles}. 
Fig. \ref{fig:platooning_notation} provides a visual representation of a homogeneous vehicle platoon {adopting a simple predecessor following topology} with arrows indicating the flow of information through the wireless communication network.  


\subsection{Platooning Topologies and Spacing Policy}

\ND{Platoons can possess distinct features out of which two crucial features are the information flow topology (IFT) and the spacing policy. }

\subsubsection{Information flow topology}
\ND{
A vehicle platoon includes a lead vehicle and $N$ follower vehicles, i.e. a total of $N+1$ vehicles. Vehicles in a platoon communicate with each other using various information flow topologies. In the information exchanged between the vehicles can be one or more states of vehicle systems, like position, velocity, and acceleration. Some common information flow topologies include \cite{Zheng2014InfluenceFormation}  \TC{Predecessor Following (PF), Predecessor-Leader Following (PLF), Bidirectional (BD), Bidirectional-Leader (BDL), Two Predecessors Following (TPF), Two Predecessor-Leader Following (TPLF), and Multiple-Predecessor Following (MPF).}}
\ND{In this work, we have demonstrated results from experiments using the MPF topology. 
}



The MPF technique is a platooning strategy that integrates information from several preceding vehicles in the control loop, starting with the immediate predecessor. 
By leveraging weighted data from the leading segment of a vehicle platoon, this method enhances performance during braking, merging, or dividing the platoon \cite{Flores2018ControlCar-following}.
This strategy proves advantageous in situations where disruptions within the platoon might occur, such as a pedestrian crossing or a collision involving a platoon member vehicle. It also enables vehicles upstream to react promptly to disturbances downstream, even before they are perceived by the immediate predecessor.

Ploeg et al. \cite{Ploeg2014LpPlatooning} have examined the two-vehicle look-ahead approach and compared it with the predecessor following strategy. In the two-vehicle look-ahead approach, the controller of the $i$-th vehicle incorporates the state of both the $(i-1)$-th and $(i-2)$-th vehicles to calculate the motor throttle command. This approach has demonstrated commendable string stability results.

Hence, this work focuses on the MPF topology approach, and subsequent sections present experiments showcasing its benefits.

\subsubsection{Spacing Policy}
\ND{Spacing policy gains paramount significance in the domain of Vehicle Platooning and CACC systems. In platooning, a group of vehicles travel closely together, benefiting from reduced aerodynamic drag and increased fuel efficiency. This necessitates a meticulously defined spacing policy that ensures a precise yet safe separation between each vehicle, allowing for synchronized movement while mitigating the risk of collisions. In CACC systems, vehicles communicate with one another, dynamically adjusting their speed and distance based on real-time data exchange. This demands a spacing policy, which is fine-tuned to account for the rapid adjustments in speed and position.}

\ND{This work employs the time delay spacing policy, a crucial element in optimizing the operation of Vehicle Platooning and CACC systems. The time delay spacing policy considers not only the physical distance between vehicles but also the temporal gap between them. By factoring in the system's response time, it accounts for the time it takes for a vehicle to detect a change in the lead vehicle's behavior and respond accordingly. This temporal dimension is particularly significant in CACC systems where vehicles rely on instantaneous data exchange for coordinated movement. The time delay spacing policy ensures that there is sufficient reaction time to adjust speed and maintain a safe separation, thus preventing unnecessary braking and acceleration cycles.}


\subsection{System Design of CACC}\label{AA}
Advancements in automobile technology have facilitated the widespread adoption of Adaptive Cruise Control (ACC) in vehicles, spurring research into autonomous car-following systems. Further refining this system, a more automated system known as CACC, leverages the improved sensor capabilities of modern vehicles. Thus, this is the system utilized for implementing connected vehicle platooning.
\begin{figure}[h]
    \centering
    \includegraphics[width=0.48\textwidth]{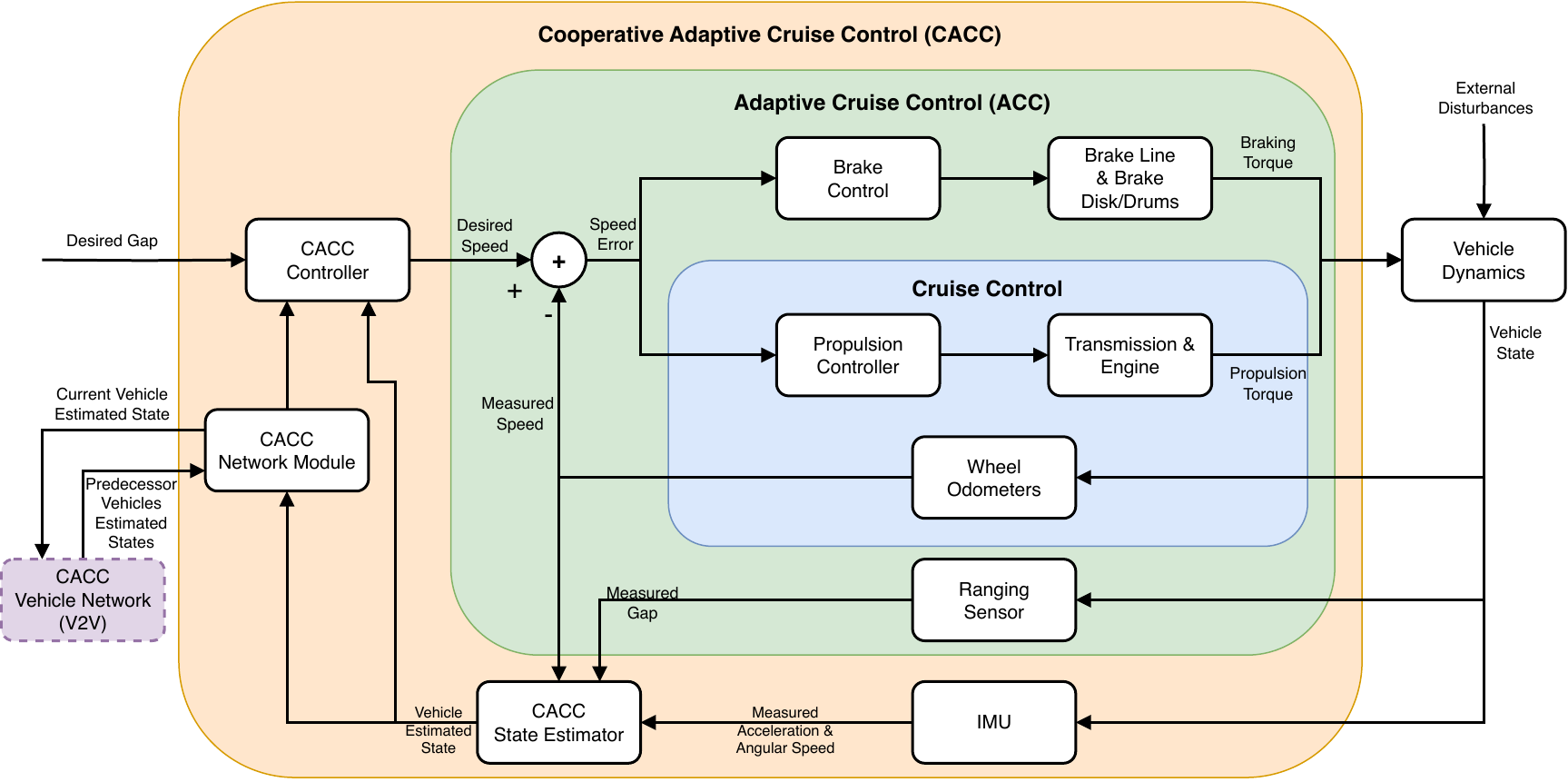}
    \caption{Control structure of a CACC system. 
    }
    \label{fig:cacc_system}
\end{figure}

Fig. \ref{fig:cacc_system} depicts the control structure of a typical CACC system, which utilizes vehicle-to-vehicle (V2V) communication to exchange critical data on position, velocity, and acceleration between equipped vehicles \cite{Gao2015AdaptiveVehicles}. This communication enables faster response times in situations where leading vehicles experience speed fluctuations. The primary objectives of CACC, as outlined in \cite{Shladover2014UsingStreams}, are:
\begin{itemize}
    \item Increase traffic throughput and maintain stable vehicle following by allowing shorter distances between vehicles in the same lane.
    \item Reduce the risk of rear-end collisions by improving communication and response times for vehicle state changes.
    \item Enhance driver comfort by minimizing propagated disturbances and functioning as an Advanced Driver Assistance System (ADAS).
    \item Potentially reduce fuel consumption by optimizing traffic flow and enabling aerodynamic drafting when distances are sufficiently close.
\end{itemize}

This work uses the system design of the CACC system as a reference to implement an experimental vehicle platform system that achieves the objectives of CACC and demonstrates a vehicle platoon.

\subsection{Objectives}
The motivation for this work lies in the limited study on platooning controllers using hardware demonstrations in physical space. 
This work focuses on implementing distributed controllers on multi-vehicle systems in a physical environment. These experiments examine the impact of communication limitations and environmental disturbances on homogeneous platoons. The comparison considers parameters such as the spacing policy and network topology in dynamic environments. In addition, scenarios are assessed to ensure controller stability for varying platoon leader velocities.
Other influences like wireless network topologies and environmental disturbances are not within the scope of this work.

\ND{This work also aims to develop both hardware and software frameworks for conducting vehicle platoon experiments. These frameworks should accurately replicate the behavior of full-sized vehicle platoons, facilitating a seamless progression from theoretical models to simulations and ultimately to tangible real-world hardware. This approach accelerates the refinement process of the platooning controller development, enabling quicker development prior to undertaking comprehensive testing on large-scale vehicles.}


\section{Problem Description}
\label{sec:problem_description}

\subsection{Vehicle Model}

This study employs a vehicle model that has been referenced from various prior research works, such as \cite{Ploeg2014LpPlatooning}. Within a platoon comprising of N vehicles, the vehicle model is defined by the following set of equations:

\begin{equation}
\label{eq:vehicel_model}
\begin{cases}
\dot{p}_{i}(t) = v_{i}(t), \\
\dot{v}_{i}(t) = a_{i}(t), \\
\tau_{i}\dot{a}_{i}(t) + a_{i}(t) = u_{i}(t), \\
\end{cases}
\end{equation}
where, for the $i$th vehicle, $p_{i}(t)$ represents position, $v_{i}(t)$ signifies velocity, $a_{i}(t)$ denotes acceleration, and $u_{i}(t)$ is the control input. Additionally, $\tau_i > 0$ represents the time-lag in the powertrain. In scenarios involving multiple predecessors, preceding vehicles relay their state information through wireless communication to vehicle $i$. Fig. \ref{fig:platooning_notation} illustrates the notation for both predecessor and successor vehicles connected in this network. The desired distance between vehicle $i$ and the $l$th vehicle ahead of it is expressed as follows \cite{Bian2019ReducingCommunication}:
\begin{equation}
\label{eq:desired_distance}
d_{i,i-l}(t) = \sum^i_{k=i-l+1} (h_k v_k(t) + d_k),
\end{equation}
where $h_k \ge 0$ stands for the time-headway of the $k$th vehicle, and $d_k > 0$ represents the anticipated standstill distance from the rear of vehicle~$k-1$ to the front of vehicle~$k$.

\subsection{Controller Design}

A linear feedback controller is employed to calculate the control signal input to the vehicle at the $i${th} position \cite{Bian2019ReducingCommunication}.
\ND{This controller is equipped with the ability to access the states of its preceding vehicles through a localized wireless communication network established among the platoon's vehicles. Consequently, it becomes possible to compare the differences between its own states and those of all its predecessors. It is assumed that the wireless communication introduces a uniform time delay denoted by $\Delta$. 
}
Therefore, the following control law for vehicle $i$ is introduced  in \cite{Abolfazli2021OnDelays}:
\begin{equation}
\begin{split}
  u_{i}(t) = &\sum^{r_i}_{l=1} \Big[ k_{pi} \big\{ p_{i}(t - \Delta) - p_{i-l}(t - \Delta) \\
   &+ \sum^{i}_{k=i-l+1}(h_k v_k(t - \Delta) + d_k) \big\}  \\
   &+ k_{vi} \big\{v_{i}(t - \Delta) - v_{i-l}(t - \Delta) \big\} \\
   &+ k_{ai} \big\{a_{i}(t - \Delta) - a_{i-l}(t - \Delta) \big\} \Big],
\end{split}
\label{eq:control_input_with_uniform_delay}
\end{equation}
where $r_i$ denotes the number of vehicles preceding vehicle~$i$ that send their information to it, with $r_i \le i$, \EA{and $k_{pi}$, $k_{vi}$, and $k_{ai}$ are tunable control gains.}

\subsection{Stability Analysis}
The goal in designing a controller for a vehicular platoon is to ensure both individual node stability, termed as \textit{internal stability}, and the overall stability of the platoon, referred to as \textit{string stability}. In this section, we provide summaries of the stability analysis studied by Abolfazli \textit{et al.} \cite{Abolfazli2021OnDelays}.
\subsubsection{Internal Stability}
Internal stability implies that the closed loop system remains stable, enabling vehicles in the platoon to track the desired inter-vehicle distance and maintain the target velocity. For internal stability to be achieved mathematically, the conditions below must be met for $1 \le l \le r$,
\begin{equation}
\label{eq_vehicel_model}
    \begin{cases}
       \lim_{t \to\infty} \big(  p_{i}(t) - p_{i-l}(t) + d_{i,i-l}(t) \big) = 0, \\
      \lim_{t \to\infty} \big( v_{i}(t) - v_{i-1}(t) \big) = 0, \\
      \lim_{t \to\infty} \big(a_{i}(t) - a_{i-l}(t) \big) = 0.
    \end{cases}  
\end{equation}
Based on Theorem 1 from \cite{Abolfazli2021OnDelays} that outlines the sufficient condition for internal stability, when the control gains~($k_{pi}$, $k_{vi}$, $k_{ai}$)  are chosen such that the following conditions hold
\begin{subequations}
\begin{align}
\begin{split}
k_{pi} &> 0, \\
\end{split}\\
\begin{split}
k_{ai} &> 0, \\
\end{split}\\
\begin{split}
k_{pi} - \tau_i (k_{vi} + k_{pi} h_i ) + \tau_i^2 k_{pi} &\neq 0, \\
\end{split}\\
\begin{split}
k_{vi} + k_{pi} h_i )  \geq k_{pi} \tau_i,
\end{split}
\end{align}
\label{eq:internal_stability_conditions}
\end{subequations}
\noindent the closed loop system is asymptotically stable for any time delay $\Delta$ that satisfies the following inequality
\begin{equation}
\label{eq:internal_stability_specification}
    \Delta r_i ( k_{vi} + k_{pi} h_i ) < 1, i \in \mathbb{N}.
\end{equation}

\subsubsection{String Stability}
Beyond internal stability, string stability is essential for vehicle platoons. They are considered string stable when disturbances do not amplify along the vehicle string. Conversely, string instability can result in traffic congestion and a greater chance of rear-end collisions.
In the work presented in \cite{Abolfazli2021OnDelays}, string stability is investigated for homogeneous platoons, where all vehicles are identical. This leads to simplified parameters such as ${\tau_i=\tau}, {r_i=r}, {h_i = h}, {k_{pi} = k_p}, {k_{vi} = k_v}, {k_{ai} = k_a}$. Following string stability functions are derived in \cite{Abolfazli2021OnDelays}:
\begin{equation}
    \label{eq:transfer_function_H}
\begin{split}
&H_{l}(s)=\\
&\frac{k_{a} s^{2}e^{-\Delta s}+\left(k_{v}-k_{p} h(r-l)\right) s e^{-\Delta s}+k_{p}e^{-\Delta s}}
{\tau s^{3}+s^{2}+r k_{a} s^{2}e^{-\Delta s}+r\left(k_{v}+k_{p} h\right) s e^{-\Delta s}+r k_{p}  e^{-\Delta s}}.
\end{split}
\end{equation}
String stability of the platoon can be guaranteed if, \cite{Bian2019ReducingCommunication}
\begin{equation}
\label{eq:string_stability_specification}
\left\|H_{l}(j \omega)\right\|_{\infty} \leq \frac{1}{r},\quad \forall 1 \leq l \leq r ,
\end{equation}
where $H_l(j\omega)$ can be derived from \eqref{eq:transfer_function_H} by substituting {${s=j\omega}$}. According to Theorem 2 from \cite{Abolfazli2021OnDelays},  system \eqref{eq:vehicel_model} with the control input \eqref{eq:control_input_with_uniform_delay}, is string stable, i.e., the string stability speciﬁcation \eqref{eq:string_stability_specification}
holds, when internally stable and when all  the following conditions are satisfied:
\begin{subequations}
	\label{eq:string_stability_conditions}
	\begin{align}	
	&  {k_{v} + k_{p}(h-\tau) \geq 0}, \label{ss region theorem - 1} \\
	&  {2\tau\Delta - \Delta h - \tau h \leq 0}, \label{ss region theorem - 2} \\
	&	k_{a} - \tau (k_v+k_ph) \leq 0,\label{ss region theorem - 3} \\
	&	\tau - 2 r k_a \Delta \geq 0, \label{ss region theorem - 4}\\ 
	&	1 + 2 r \Big(k_{a} - \tau (k_v+k_ph) \Big) + 2 r \Delta \Big(k_p (\tau - h) - k_v \Big) \geq 0, \label{ss region theorem - 5} \\ 
	&	r^2 k_p^2 h^2 (1-(r-l)^2) + 2 r^2 k_p k_v h (1+r-l) \nonumber\\
	& \qquad \qquad \qquad  \qquad \qquad - 2 r k_p \geq 0, \quad 1\leq l \leq r \label{ss region theorem - 6}.
	\end{align}
\end{subequations}
For the region defined by equation~\eqref{eq:string_stability_conditions}, there exists a set of feedback gains $k_p$, $k_v$ and $k_a$, such that string stability holds if,
\begin{equation}
\label{eq:hmin_criteria}
h \geq h_{min} = \frac{ 2(\tau + \Delta) }{ 2 r k_a + 1 }.
\end{equation}

\section{System Architecture}
\label{sec:system_architecture}


\subsection{Architecture}


\subsubsection{Experiment Platform - Hardware}
\label{sec:Experiments_ExperimentPlatform_Hardware}
This project aims to enhance platooning algorithms for autonomous vehicles. To replicate real-world platooning, the experimental platform must meet specific criteria \cite{SamuelMitchell2016GROUNDENVIRONMENT}:

\begin{list4}
    \item[1.] The platform's dynamics, particularly in the longitudinal direction, should resemble a full-scale vehicle.
    \item[2.] It must reach speeds required for small scale platooning demonstrations.
    \item[3.] In case of controller failure, the robot should withstand collisions.
\end{list4}

The chosen Ubiquity Robotics' Magni robot platform meets these criteria. Despite being a differential drive robot, its longitudinal dynamics closely resemble those of a full-scale vehicle. It can achieve speeds of up to 1m/s (or higher with updates), sufficient for experimental simulations in the lab environment.
\begin{figure}[t]
    \centering
    \includegraphics[width=0.46\textwidth]{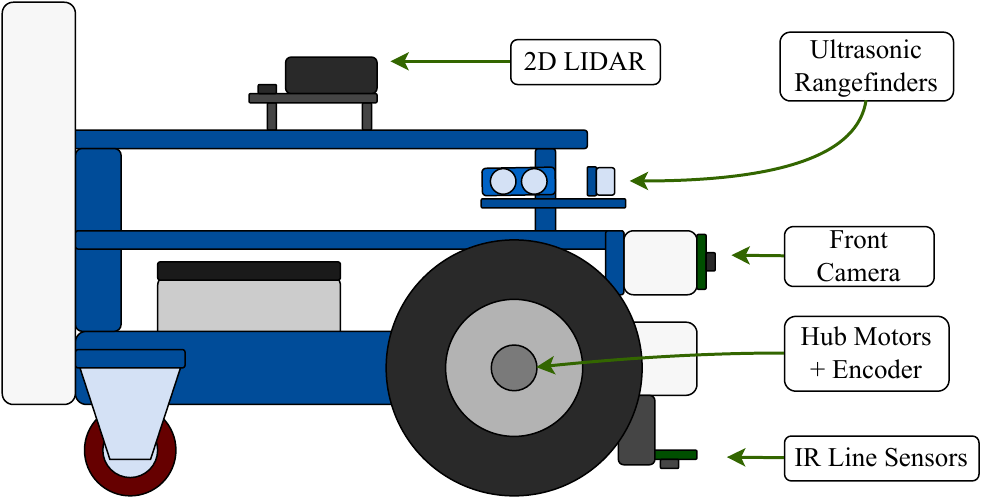}
    \caption{A side view of the Magni robot with sensors and actuators.}
    \label{fig:magni_robot}
\end{figure}

\subsubsection{Experiment Platform - Software Framework}
The software system for the experimental platform must meet specific requirements to effectively support the experimental setup. Key requirements include:

\begin{list4}
    \item Seamless integration of various CACC controllers with minimal configuration effort.
    \item Capability to record sensor data and control commands.
    \item Minimization of processing overhead and other software application impacts on controller performance, especially in the presence of significant delays.
    \item Modularity to facilitate prototyping and future sub-component development.
\end{list4}

To address these requirements, the Robot Operating System (ROS) was chosen as the framework for software architecture implementation. ROS, an open-source middleware system supporting C++ and Python modules, proves highly advantageous for prototyping experimental platforms. It facilitates message passing, task scheduling, data logging, and offers flexibility in hardware selection. ROS aligns well with the needs of the experimental platform and was thus selected as the software framework for this system. 

In Rajamani's work \cite{rajamani2011vehicle}, a split-controller configuration is discussed for automotive applications. This involves a high-level controller handling linear and angular velocity/acceleration commands, while a low-level controller executes these commands on the motors. For effective command relay, the low-level controller must operate at a significantly higher rate (approximately 50 Hz) compared to the high-level controller (around 10 Hz).

The system comprises two main components: the experimental platform vehicle and the ground control station (GCS). Detailed descriptions of the software architecture and its various modules follow in subsequent subsections. The platform vehicle simulates the function of a real world vehicle hosting the controller under test. The GCS is a remote workstation which enables the monitoring and diagnostics of the experimental setup in a laboratory environment.

\begin{figure}[h]
    \centering
    \includegraphics[width=0.48\textwidth]{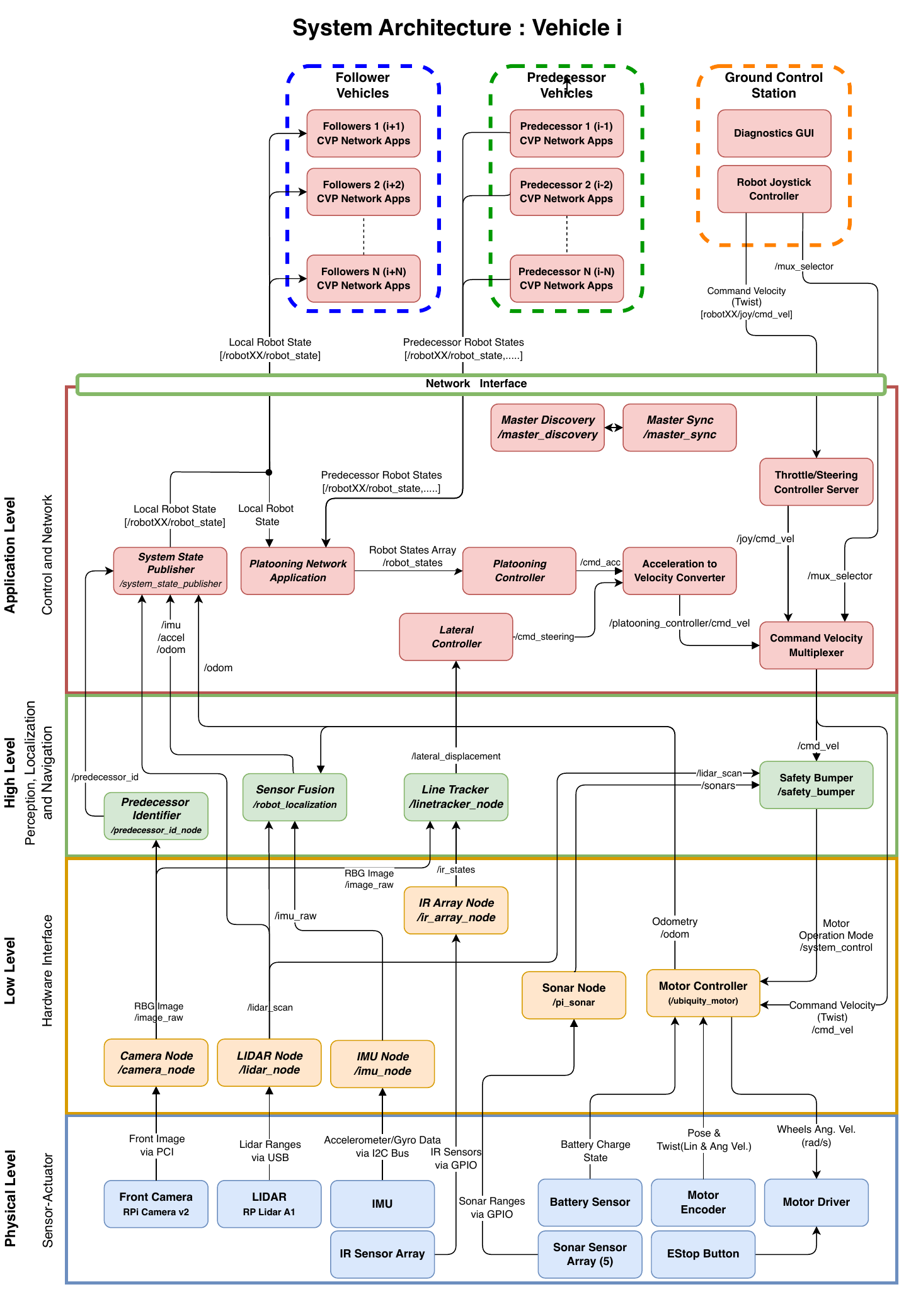}
    \caption{Diagram of System Architecture for the ROS-based software system on the experiment platform.
    }
    \label{fig:system_achitecture_diagram}
\end{figure}

Fig. \ref{fig:system_achitecture_diagram} provides an overview of the system architecture, showcasing the software running on the onboard computer within each experimental platform vehicle. Additionally, it displays the network interface modules responsible for data exchange with other vehicles in the platoon and the GCS. The architecture is designed for a ROS system, where the arrows between the subsystem modules indicate the ROS topics or services being published and subscribed by the respective modules.

\section{Experiments}
\label{sec:experiments}


\subsection{Experiment Setup}
\label{sec:Experiment_HardwareSetup_ExperimentSetup}

\begin{figure}[h]
    \centering
    \includegraphics[width=0.48\textwidth]{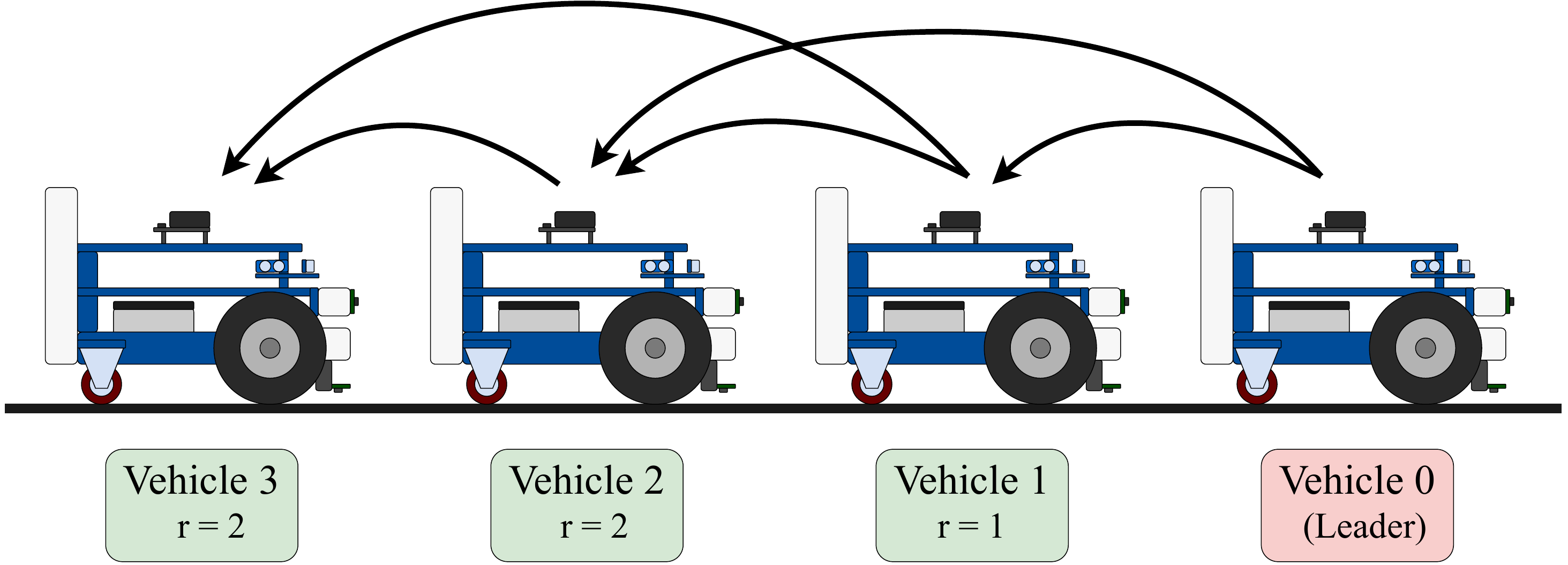}
    \caption{Two-Predecessor Following - Experiment Setup.}
    \label{fig:two_pf_expt_setup}
\end{figure}

In the platooning experiment, the goal is to assess the performance of the controller in \cite{Abolfazli2021OnDelays} developed to operate under the  Multiple Predecessors Following (MPF) topology. The experiment employs custom-designed experimental platform vehicles detailed in section \ref{sec:Experiments_ExperimentPlatform_Hardware}. The setup involves a lead vehicle followed by identical vehicles in the sequence: Lead Vehicle, Vehicle 1, Vehicle 2, Vehicle 3, and so forth. 

Fig. \ref{fig:two_pf_expt_setup} illustrates the configuration of these experimental vehicles and the data flow direction according to the MPF methodology. Each vehicle is equipped with a longitudinal controller, which is described below:

\begin{list4}
    \item \textbf{Leader Vehicle}: The controller on the lead vehicle executes a predefined acceleration control input command, specifically $0.1 m/s^2$ acceleration at $t=5 s$. At $t=15 s$, a brief disturbance deceleration is introduced, affecting the leader vehicle's throttle input with ${u_0 = A_d \sin(\omega_0 t+\pi)}$ for half a cycle, i.e., $\frac{\pi}{\omega_0}$ seconds. To bring the platoon to a stop, a constant braking deceleration is applied at $t=40 s$. Fig. \ref{fig:leader_input_command} visualizes the leader controller's throttle (acceleration) input command.
\begin{figure}[b]
    \centering
    \includegraphics[width=0.49\textwidth]{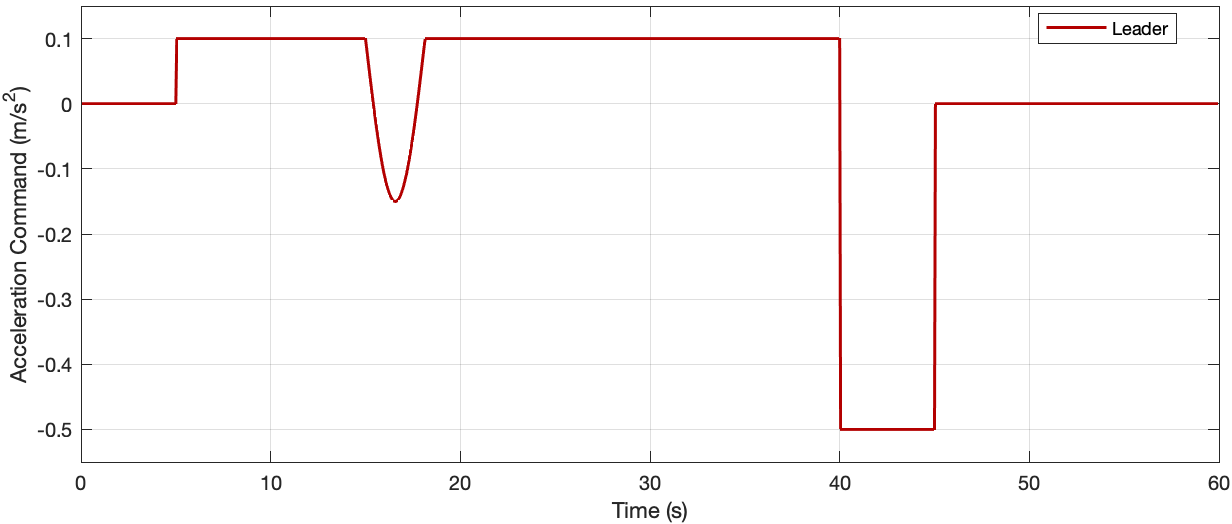}
    \caption{Leader Vehicle : Acceleration Input Command.}
    \label{fig:leader_input_command}
\end{figure}
\item \textbf{Vehicle 1}: The controller on this vehicle employs the MPF controller described in equation (\ref{eq:mpf_controller_vehicle1}) with $r = 1$, where $u_1(t) = u_1 $, $h = h_1$, $d = d_1$, $k_p = k_{p1}$, $k_v = k_{v1}$ and $k_a = k_{a1}$.

    \begin{equation}
    \label{eq:mpf_controller_vehicle1}
    \begin{split}
    u_1 = &- k_{p}( p_1 - p_0 + h v_1 + d) \\
            &- k_{v} ( v_1 - v_0) - k_{a} (a_1 - a_0).
    \end{split}
    \end{equation}
    
    \item \textbf{Vehicle 2, Vehicle3, ...}: This vehicle's controller utilizes the MPF controller described in equation (\ref{eq:mpf_controller_vehicle2}) with $r = 2$, where $u_2(t) = u_2$,  $h = h_1 = h_2$, $d = d_1 = d_2$, $k_p = k_{p2}$, $k_v = k_{v2}$ and  $k_a = k_{a2}$.
    \begin{equation}
    \begin{split}
    \label{eq:mpf_controller_vehicle2}
    u_i = &- k_{p}[( p_i - p_{i-1} + h v_i + d) \\
        &+ ( p_i - p_{i-2} + h v_i + d + h v_{i-1} + d) ] \\ 
        &- k_{v} [( v_i - v_{i-1}) + ( v_i - v_{i-2})] \\
        &- k_{a} [(a_i - a_{i-1}) + (a_i - a_{i-2})]. \\
    \end{split}
    \end{equation}   
    
    
\end{list4}

\subsection{Results}

The experiment involved a four-vehicle platoon, and the setup, as illustrated in Fig. \ref{fig:two_pf_expt_setup} and detailed in section \ref{sec:Experiment_HardwareSetup_ExperimentSetup}, was employed. This experiment assesses the TPF controller, where each vehicle's controller is designed with $r\le2$. The specific parameter values used for this experiment are listed in Table \ref{table:two_pf_parameters}.

The parameters $(k_p, k_v, k_a)$ are set to $(0.1, 0.61, 0.4)$ to satisfy the condition in \eqref{eq:internal_stability_conditions}. Values for $\tau = 0.9$ and $\Delta=0.05$ are determined through measurements on the experimental platform vehicle. With these parameters and $r=2$, both the internal stability specification \eqref{eq:internal_stability_specification} and the string stability criteria \eqref{eq:string_stability_conditions} are met. The minimum time headway for these parameters is $h_{min} = 0.719$, and the chosen value $h=0.78$ exceeds $h_{min}$, satisfying \eqref{eq:hmin_criteria}.
\begin{table}[t]
\caption{Experiment Parameters for Two Predecessor Following}
\label{table:two_pf_parameters}
\begin{tabular}{lrl}
\hline
\textbf{Parameter} & \multicolumn{1}{l}{\textbf{Value}} & \textbf{Description}                                                         \\ \hline
N                  & 4                                  & Total Number of Vehicles in the Platoon                                      \\
r                  & 2                                  & Number of predecessor vehicles                                               \\
$\tau$             & 0.9                                & Timelag in the power train, $\tau > 0$                               \\
h                  & 0.78                               & Time headway, $h > 0$                                                \\
d                  & 0.6                                & Desired standstill gap, $d > 0$                                       \\
$k_p$              & 0.1                                & Tunable gain for feedback distance                                           \\
$k_v$              & 0.61                               & Tunable gain for feedback velocity                                           \\
$k_a$              & 0.41                               & Tunable gain for feedback acceleration                                       \\
$\Delta$           & 0.05                               & Time delay                                                      \\
$A_0$                 & 0.05                               & Amplitude of throttle to leader vehicle's controller                          \\
$A_{d}$                & 0.25                               & Amplitude of disturbance                                                     \\ \hline
\end{tabular}
\end{table}

\subsubsection{Simulation Results}
\label{sec:ExperimentalEvaluation_TwoPF_SimulationResults}
This section presents the results obtained from running the Simulink model with a mathematical model simulation of the experiment described in .

\begin{figure}[t]
    \centering
    \includegraphics[width=0.49\textwidth]{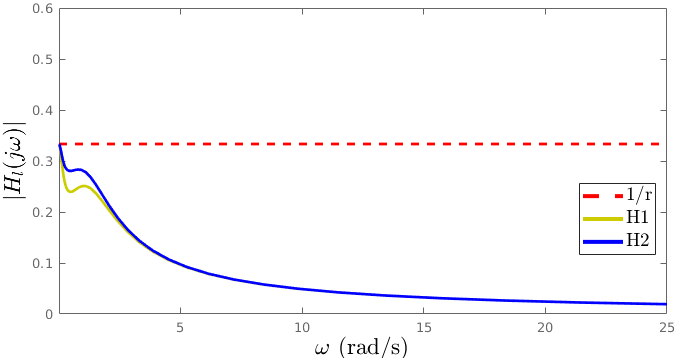}
    \caption{Simulation Results for Two-Predecessor Following - String Stability.}
    \label{fig:two_pf_sim_string_stability}
\end{figure}

For stability analysis, Fig. \ref{fig:two_pf_sim_string_stability} displays the magnitude-frequency diagram of the function $|H_i(j \omega)|$. In each case, $|H_1(j \omega)|$ and $|H_2(j \omega)|$ (for $l=1$ and $l=2$, respectively) do not exceed the maximum tolerable value for string stability, i.e., $\frac{1}{r}$ according to \eqref{eq:string_stability_specification}. This confirms the satisfaction of \eqref{eq:string_stability_specification} and verifies the string stability of the platoon.

\begin{figure}[t]
    \centering
    \includegraphics[width=0.49\textwidth]{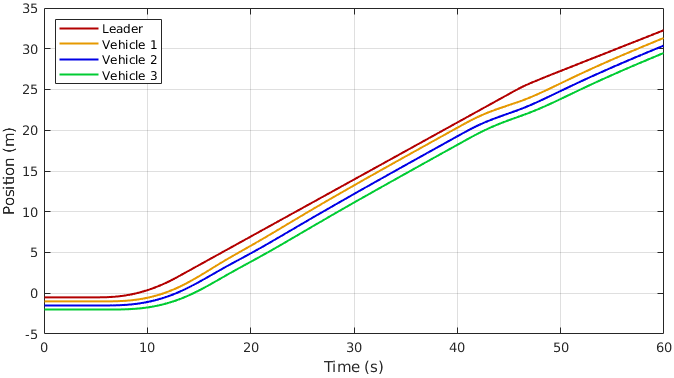}
    \caption{Simulation Results for Two-Predecessor Following - Distance travelled by each vehicle\vspace{-0.2cm}.}
    \label{fig:two_pf_sim_distance_travelled}
\end{figure}

\begin{figure}[h]
    \centering
    \includegraphics[width=0.49\textwidth]{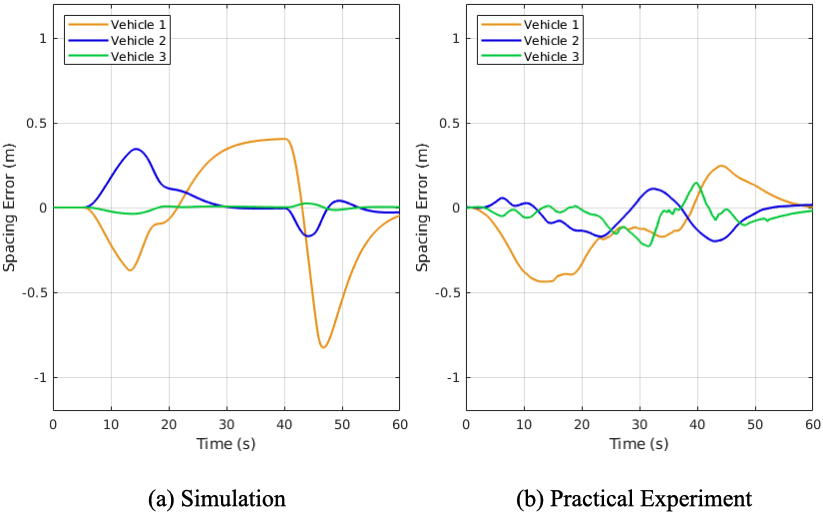}
    \caption{Comparison of the results of simulation and practical experiment for the Two-Predecessor Following - Spacing Errors.}
    \label{fig:two_pf_spacing_error}
\end{figure}

Next, Fig. \ref{fig:two_pf_sim_distance_travelled} illustrates the absolute distances covered by each vehicle in the platoon after simulating for 60 seconds. The platoon vehicles start from a stationary position, and the platoon leader vehicle initiates acceleration at $t=5$ seconds. At $t=15$ seconds, a disturbance in the form of deceleration is introduced to the leader vehicle's throttle input, causing perturbations in the platoon's longitudinal direction. Fig. \ref{fig:two_pf_spacing_error}(a) illustrates the spacing error of the vehicles relative to their predecessor vehicles.

\begin{figure}[h]
    \centering
    \includegraphics[width=0.49\textwidth]{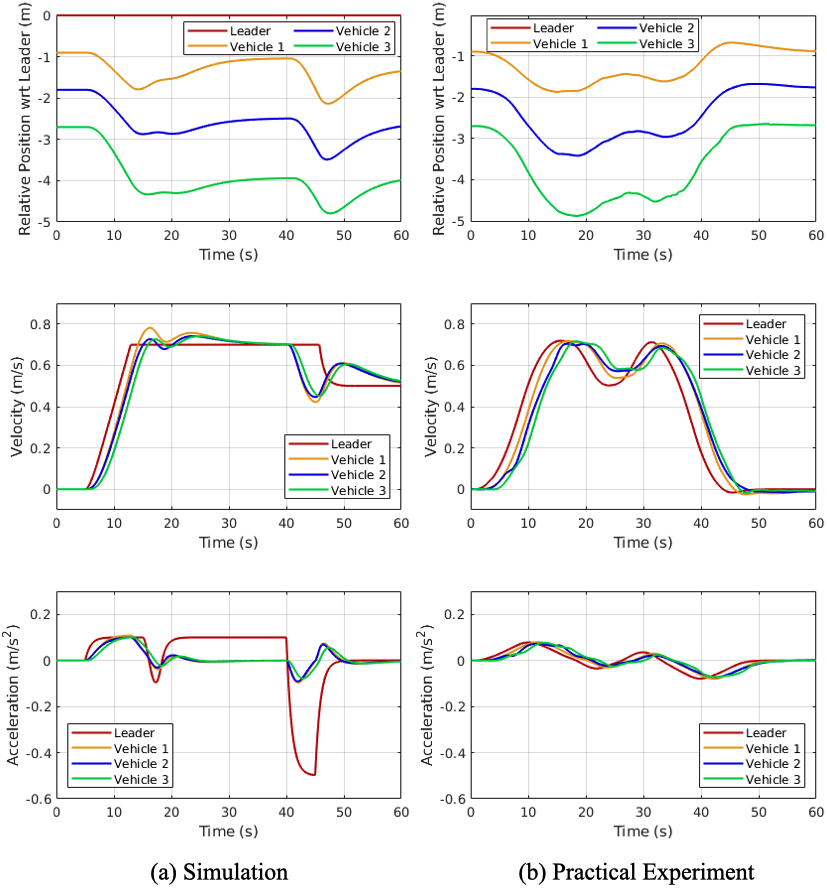}
    \caption{Comparison of the results of simulation and practical experiment for the Two-Predecessor Following - System States.}
    \label{fig:two_pf_states}
\end{figure}

The states (relative position with respect to the leader vehicle, velocity, and acceleration) of the simulated platoon vehicles are presented in Fig. \ref{fig:two_pf_states}(a). Similarly, Fig. \ref{fig:two_pf_spacing_error}(a) showcases the spacing errors of the follower vehicles based on data from the simulation.

\subsubsection{Test Environment and Experiment Results}

In this section, we delve into the results of the practical experiment conducted on the experimental platform vehicle\footnote{\TC{A video in which the experimental setup and the operation of the vehicle platoon system are demonstrated can be found in \url{https://www.youtube.com/watch?v=-r5EvBTVL1Q}}.}.

The outcomes of this experiment, depicted in Fig. \ref{fig:two_pf_states}(b), provide insights into various aspects. These include the relative positions, velocities, and accelerations of each vehicle in relation to the platoon leader.

To facilitate data acquisition, a LIDAR sensor was strategically positioned at the center of the experimental vehicle's roof, precisely 0.2 meters from its front. This positioning ensures that the plotted relative positions reflect the distance from the rear of the leader vehicle to the center of the $i$-th vehicle.

In alignment with the simulation setup, the experiment commenced with the platoon vehicles at rest. Subsequently, the leader vehicle initiated acceleration at $t=5$ seconds. Similarly, a disturbance was introduced as a deceleration to the leader vehicle's throttle input at $t=15$ seconds. This deliberate perturbation was designed to simulate real-world scenarios, introducing longitudinal perturbations to the platoon. The resulting spacing errors of the vehicles in relation to their predecessor vehicles are vividly illustrated in Fig. \ref{fig:two_pf_spacing_error}(b).

\subsection{Discussion}
This section conducts a comparative analysis of the obtained results.

Fig. \ref{fig:two_pf_sim_string_stability} demonstrates that the platoon adheres to the conditions of string stability for the selected parameter configurations in both the two-predecessor following and three-predecessor following platoon experiments. These results align with the findings of Abolfazli \textit{et al.} \cite{Abolfazli2021OnDelays}.

In the experiments conducted with the experimental platform vehicle, Fig. \ref{fig:two_pf_states}  showcase that the vehicle controllers effectively generate the necessary acceleration commands to trail the predecessor vehicle, mirroring the simulation outcomes. When comparing the spacing errors in the practical experiment to the simulation results (as depicted in Fig. \ref{fig:two_pf_spacing_error}), it becomes apparent that the anticipated spacing error doesn't perfectly replicate from the simulation to the real-world scenario. This emphasizes that while the system maintains internal stability and string stability, the efficacy of the MPF controller is influenced by various unaccounted factors in its control input equation.

The discrepancies in spacing may be attributed to physical interactions with the environment, which may encompass the following factors:

\begin{list4}
    \item \textbf{Sensor Fusion and Measurement Errors}: All sensors, including LIDARs, IMUs, and odometers employed in the vehicle designed for this experiment, have inherent error margins. While these errors are factored into the sensor fusion model, they can lead to minor inaccuracies over time.
    \item \textbf{Friction}: The interaction with the experiment area's floor may induce uneven friction on each wheel. This could result in differential slipping of wheels, causing the vehicle to turn instead of moving forward, due to the distinctive dynamics of the experimental vehicle platform.
    \item \textbf{Communication}: The vehicles in this experiment communicate over a wireless LAN network utilizing the TCP/IP communication protocol. Transmission over a wireless medium is subject to substantial noise and delays, subsequently leading to additional delays in processing the controller's throttle command in each follower vehicle.
\end{list4}

These experimental outcomes present a valuable opportunity for scrutinizing the MPF controller's performance in a real-world environment, offering avenues for refinement based on external environmental feedback.



\section{Conclusions and Future Directions}
\label{sec:conclusions_futuredirections}
\subsection{Conclusions}
\label{sec:Conclusions}

In summary, the experiments conducted in this work demonstrate that the controller under investigation effectively maintains stable vehicle following in both simulations and real-world scenarios for the MPF topology, despite some spacing variations caused by environmental factors. These findings suggest opportunities for further enhancement, particularly in addressing sensor inaccuracies, frictional forces, and time-varying communication delays to improve the controller’s reliability in practical cases.


\subsection{Future Directions}
\label{sec:Future_Directions}
Ensuring safety is imperative, particularly given the deployment of CACC on public roads alongside heavy vehicles \cite{shladover2018using}. The close proximity of vehicles in platooning scenarios raises the potential for collisions within the platoon and with surrounding vehicles, especially in the event of disruptions or malfunctions. Hazards may arise from external factors like sudden maneuvers from neighboring vehicles, adverse road conditions, and inclement weather. The forward-facing range sensor, typically a radar, plays a pivotal role for CACC controllers as it measures inter-vehicle distances and relative speeds. Research on implementing redundant systems for these measurements can enhance accuracy and serve as a fail-safe in the event of internal subsystem failures \cite{Axelsson2017SafetyReview}.

Wireless communication is vital for CACC controllers to share up-to-date information from other platoon vehicles for real-time operation. The study's results demonstrate increased packet losses and delays in the local network when the number of platooning vehicles exceeds a certain number (in this case, three). Investigating the impact of these losses and delays in the communication network would be valuable \cite{Aramrattana2018APlatooning}. The current software architecture employs ROS as a communication framework, but there is room for improvement in information exchange efficiency. Upgrading the networking subsystem within ROS to employ more efficient protocols, coupled with standardized V2V communication protocols and optimized IFTs, could significantly enhance information transfer reliability in this experimental setup \cite{Wang2019CooperativeConstraints}.


\bibliographystyle{ieeetr} 
\bibliography{ref_cacc_ecc24}

\end{document}